\documentclass[twocolumn,showpacs,preprintnumbers,amsmath,amssymb]{revtex4}

\usepackage{graphicx}
\usepackage{dcolumn}
\usepackage{bm}

\usepackage{amssymb,bm,color}

\def\eqref#1{Eq.~(\ref{#1})}

\def\Eq#1{\begin{equation} #1 \end{equation}}
\def\Eqr#1{\begin{eqnarray} #1 \end{eqnarray}}
\def\Eqrsubl#1#2{\begin{subequations}\label{#1}\Eqr{#2}\end{subequations}}

\newcommand{\nn}{\nonumber}
\newcommand{\pd}{\partial}
\newcommand{\bea}{\begin{eqnarray}}
\newcommand{\eea}{\end{eqnarray}}

\def\Xsp{{\rm X}}

\def\Ysp{{\rm Y}}
\def\Zsp{{\rm Z}}
\def\X5sp{{\rm X}_5}
\def\Y3sp{{\rm Y}_3}
\def\Z3sp{{\rm Z}_3}

\def\lap{{\triangle}}
\def\e{{\rm e}}

\begin{document}

\preprint{YITP-11-101}

\title{
Warped de Sitter compactifications 
in the scalar-tensor theory
}

\author{Masato Minamitsuji}
\affiliation{
Yukawa Institute for Theoretical Physics,
Kyoto University, Kyoto 606-8502, Japan.
}%

\author{Kunihito Uzawa}
\affiliation{%
Department of Physics, Kinki University,
Higashi-Osaka, Osaka 577-8502, Japan
}%

\date{\today}

\begin{abstract}
We present new solutions of warped compactifications 
in the higher-dimensional gravity coupled to the 
 scalar and the form field strengths.
These solutions are constructed 
in the $D$-dimensional spacetime with matter fields, with the internal space 
that has a finite volume. 
Our solutions give explicit examples
where the cosmological constant or 0-form field strength
leads to a 
de Sitter spacetime 
in arbitrary dimensions.
\end{abstract}

\maketitle



We present exact solutions of
warped de Sitter
compactifications
in the scalar-tensor theory coupled to
the form field strength.
We show that certain coupling constants
and the warped structure
\cite{
Maldacena:2000mw, Blaback:2010sj, Neupane:2010ya, Minamitsuji:2010cm, 
Minamitsuji:2011gn, Minamitsuji:2011gp}
allow accelerating expansion of our Universe 
\cite{Riess:1998cb, Perlmutter:1998np, Riess:2001gk,
tsujikawa}
in arbitrary dimensions.
In the higher-dimensional gravity theory, 
time-dependence of the hyperbolic compact internal space yields 
an accelerating universe in the four-dimensional Einstein frame 
without the warped structure 
\cite{Townsend:2003fx}.
Then, 
more general 
accelerating solutions 
without warping
in the context of higher-dimensional theory
have been studied extensively in
\cite{Townsend:2003fx, 
Ohta:2003pu, Ohta:2003ie, Chen:2003ij, Chen:2003dca, Chen:2006ia}.
A (warped) compactification to de Sitter or accelerating universe 
has also been invoked as models 
not only 
in higher-dimensional gravity but also 
in
supergravity and 
string theory \cite{Kachru:2003aw, Kachru:2003sx, Rosseel:2006fs, 
Silverstein:2007ac, Danielsson:2009ff, Caviezel:2009tu, 
Wrase:2010ew, Danielsson:2010bc, Blaback:2010sj}. 
In many of these applications the solutions of the
de Sitter spacetime with a warped compactification 
are evaluated at appropriate values of their parameters. 
At present 
several warped solutions which give the four-dimensional 
de Sitter spacetime 
have been reported recently in \cite{Neupane:2010ya,Minamitsuji:2011gp}
where one of the internal dimensions is noncompact,
but de Sitter compactification 
with a compact internal space 
has been explored much less extensively
(see e.g., \cite{Maldacena:2000mw}
for the arguments on general properties of
the warped de Sitter compactification).
The main purpose of 
the present work 
is to present an explicit example 
of such warped de Sitter compactification 
just as the starting point
for the research in this direction,
though
the solution presented in this Letter still may not be suitable
for the realistic cosmological models. 

Without relying on an infinite volume of extra dimensions, 
we can directly deduce a known construction of the de Sitter 
compactification 
in the framework of the higher-dimensional 
gravity theory. 
The strategy to find solutions 
basically follows one
 used in \cite{
Minamitsuji:2011gp}.

We also assume the same metric ansatz in \cite{Minamitsuji:2011gp},
except for the replacement of the noncompact direction
with that of S$^1$.
As we will see,
our relatively simple ansatz, however, 
will give us much 
information to reveal the essential properties
of the warped de Sitter compactification. 

In the following, we construct the de Sitter compactifications in 
the $D$-dimensional theory. 
Our construction is 
analogous to a 
previous 
approach to warped compactifications 
\cite{Minamitsuji:2010cm, Minamitsuji:2011gp}.
The starting point of the construction is that the internal space can 
be expressed as 
a compact manifold. 
In the first model, 
we introduce the cosmological 
constant and matter fields such as the scalar and gauge fields while 
keeping 
the scalar coupling constant labeling the fields fixed. 
This framework will give us 
the easiest framework for describing the de Sitter spacetime, 
compact internal space 
and their generalizations. 
In the 
second model,
because of the 
presence of 0-form field strength 
the cosmological 
constant 
may be taken to be negative 
(see e.g., \cite{Maldacena:2000mw} for the general arguments). 
This second solution is related to the massive theory 
which has been discussed with different coupling constant
in \cite{Bergshoeff:1996ui, Bergshoeff:1997ak}. 
The construction of compactification in terms of higher-dimensional 
theory can be carried through the solutions of the Einstein equations, 
leading to descriptions based on compact internal space 
in various dimensions, as we explain in details. 


Let us consider a gravitational theory with the metric $g_{MN}$,
the scalar field $\phi$, the cosmological constant $\Lambda$,
and the antisymmetric tensor field of rank $p_I$ .
The action which we consider is given by
\Eqr{
S&=&\frac{1}{2\kappa^2}\int 
\left[\left(R-2\e^{\alpha\phi}\Lambda\right)\ast{\bf 1}
-\frac{1}{2}d\phi\wedge\ast d\phi\right.\nn\\
&&\left.-\sum_I\frac{1}{2\cdot p_I!}
\e^{\alpha_I\phi}
F_{(p_I)}\wedge\ast F_{(p_I)}\right],
\label{g:action:Eq}
}
where $\kappa^2$ is the $D$-dimensional gravitational constant, and 
$\ast$ is the Hodge operator in the $D$-dimensional spacetime, 
$F_{(p_I)}$ is the $p_I$-form field strength,
and $\alpha$, $\alpha_I$ are constants.

The $D$-dimensional action \eqref{g:action:Eq} gives the 
field equations: 
\Eqrsubl{g:field equations:Eq}{
&&\hspace{-0.9cm}R_{MN}=\frac{1}{2}\pd_M\phi\pd_N\phi
+\frac{2}{D-2}\e^{\alpha\phi}\Lambda g_{MN}\nn\\
&&\hspace{-0.2cm}+\sum_I
\frac{\e^{\alpha_I\phi}}{2\cdot p_I!}\left[p_I
\left(F_{(p_I)}\right)^2_{MN}
-\frac{p_I-1}{D-2}g_{MN}F_{(p_I)}^2\right],
   \label{g:Einstein:Eq}\\
&&\hspace{-0.9cm}
d\left[\e^{\alpha_I\phi}\,\ast F_{(p_I)}\right]=0\,,
    \label{g:gauge:Eq}\\
&&\hspace{-0.9cm}d\ast d\phi-2\alpha\e^{\alpha\phi}\Lambda\ast{\bf 1}
-\sum_I\frac{\alpha_I\,\e^{\alpha_I\phi}}{2\cdot p_I!}F_{(p_I)}
\wedge\ast F_{(p_I)}=0,
   \label{g:scalar:Eq}
}
where $\left(F_{(p_I)}\right)^2_{MN}$ is defined by 
$F_{MA_1\cdots A_{p_I-1}}{F_N}^{A_1\cdots A_{p_I-1}}$\,.
To solve the field equations, we assume that the $D$-dimensional metric
takes the form
\Eq{
ds^2=\e^{2A(y)}\left[q_{\mu\nu}(\Xsp)dx^{\mu}dx^{\nu}
+u_{ij}(\Ysp)dy^idy^j\right],
 \label{g:metric:Eq}
}
where $q_{\mu\nu}(\Xsp)$ is 
the $n$-dimensional external spacetime metric which
depends only on the $n$-dimensional coordinates $x^{\mu}$,
and $u_{ij}(\Ysp)$ is the $(D-n)$-dimensional
internal space metric which
depends only on the $(D-n)$-dimensional coordinates $y^i$, 
the function $A(y)$ depends only on the coordinates $y^i$. 
We assume that $\Ysp$ is a smooth and compact internal manifold.

As the first model, 
we consider a scalar field and 
field strengths 
$F_{(n)}$ and $F_{(D-n)}$ which have nonvanishing components
only along the $\Xsp$ and $\Ysp$ directions, respectively.
The scalar field $\phi$ and
the gauge field strengths are assumed to be
\Eqrsubl{g:fields:Eq}{
\phi&=&-\frac{2}{\alpha} A\,,
  \label{g:ansatz of scalar:Eq}\\
F_{(n)}&=&f_1\,\Omega(\Xsp)\,,
  \label{g:ansatz of gauge1:Eq}\\
F_{(D-n)}&=&f_2\,\e^{(D-2)A(y)}\Omega(\Ysp)\,,
  \label{g:ansatz of gauge2:Eq}
}
where $f_1$ and $f_2$ are constants, and $\Omega(\Xsp)$, $\Omega(\Ysp)$ 
denote the volume $n$-, $(D-n)$-form,
\Eqrsubl{g:volume:Eq}{
\Omega(\Xsp)&=&\sqrt{-q}dx^0\wedge \cdots \wedge dx^{n-1}\,,\\
\Omega(\Ysp)&=&\sqrt{u}dy^1\wedge \cdots \wedge dy^{D-n}.
}
Here, $q$ and $u$ are the determinants
of the metric $q_{\mu\nu}$ and 
$u_{ij}$, respectively. 
In the following, we assume that the parameters $\alpha_{(n)}$, 
$\alpha_{(D-n)}$ are given by 
\Eq{
\alpha_{(n)}=-\alpha_{(D-n)}=-\alpha(n-1)\,.
   \label{g:coupling:Eq}
}

Let us first consider the Einstein Eqs.~(\ref{g:Einstein:Eq}).
Using the assumptions (\ref{g:metric:Eq}) and (\ref{g:fields:Eq}),
the Einstein equations are given by
\Eqrsubl{g:cEinstein:Eq}{
&&\hspace{-1cm}R_{\mu\nu}(\Xsp)
-q_{\mu\nu}\left[P+\frac{K}{2(D-2)}
-\frac{1}{2}f^2\right]=0,
 \label{g:cEinstein-mn:Eq}\\
&&\hspace{-1cm}R_{ij}(\Ysp)-(D-2)D_iD_jA
+\left(D-2-\frac{2}{
\alpha^2}\right)\pd_iA\pd_jA\nn\\
&&~~~~-u_{ij}\left[P+\frac{K}{2(D-2)}\right]=0\,,
 \label{g:cEinstein-ij:Eq}
}
where $f^2\equiv f_1^2+f_2^2$, and 
$D_i$ is the covariant derivatives with respect to
the metric $u_{ij}$, and $\triangle_{\Ysp}$ is  
the Laplace operator on the space of Y and
$R_{\mu\nu}(\Xsp)$ and $R_{ij}(\Ysp)$ are the Ricci tensors
of the metrics $q_{\mu\nu}$ and $u_{ij}$, respectively, 
and $P$, $K$ are defined as 
\Eqrsubl{g:def:Eq}{
P&=&\lap_{\Ysp}A+(D-2)u^{ij}\pd_iA\pd_jA\,,
\label{g:p:Eq}\\
K&=&4\Lambda +(n-1)f^2\,.
\label{g:k:Eq}
}

Next we consider the gauge field. 
Under the assumptions 
\eqref{g:ansatz of gauge1:Eq} and \eqref{g:ansatz of gauge2:Eq}, 
the Bianchi identities and 
the equations of motion for the
gauge fields are automatically satisfied. 
Substituting Eqs.~(\ref{g:metric:Eq}) and (\ref{g:fields:Eq}) into
Eq.~(\ref{g:scalar:Eq}), the scalar field equation gives
\Eqr{
P+\frac{\alpha^2}{4}K=0\,.
   \label{g:scalar eq:Eq}
}

Now we set 
\Eq{
\alpha^2=\frac{2}{D-2}+c\,,
    \label{g:c1:Eq}
}
where $c$ is constant. 
In terms of \eqref{g:fields:Eq} and \eqref{g:c1:Eq}, 
the field equations reduce to
\Eqrsubl{g:cEinstein2:Eq}{
&&R_{\mu\nu}(\Xsp)
+\frac{1}{2}q_{\mu\nu}\left(f^2+\frac{c}{2}K\right)=0,
 \label{g:cEinstein2-mn:Eq}\\
&&R_{ij}(\Ysp)
-(D-2)D_iD_jA
+\left(D-2-\frac{2}{
\alpha^2}\right)\pd_iA\pd_jA\nn\\
&&~~~~+\frac{c}{4}K u_{ij}=0,
 \label{g:cEinstein2-ij:Eq}\\
&&P+\frac{1}{4}\left(\frac{2}{D-2}+c\right)K=0\,.
   \label{g:scalar eq2:Eq}
}
If $c$ and $\Lambda$ satisfy  
\Eqr{
f^2+\frac{c}{2}\left[4\Lambda+(n-1)f^2\right]<0\,,
   \label{g:Lambda:Eq}
}
the solution leads to an accelerating 
expansion of the $n$-dimensional spacetime. 

Next we consider the Eqs.~(\ref{g:cEinstein-ij:Eq}) and 
(\ref{g:scalar eq2:Eq}).  
We assume that the $(D-n)$-dimensional metric $u_{ij}$ 
is given by 
\Eqr{
ds^2(\Ysp)&=&u_{ij}(\Ysp)dy^idy^j\nn\\
  &\equiv&
b_0^2
d\theta^2+w_{mn}(\Zsp)dz^{m}dz^{n}\,,
    \label{g:i-metric:Eq}
}
where 
$b_0$ is a constant with the dimension of the length
and $w_{mn}(\Zsp)$ 
is the metric of $(D-n-1)$-dimensional compact Einstein space 
depending on the coordinates $z^m$.

Using the metric (\ref{g:i-metric:Eq}), the 
Eqs.~(\ref{g:cEinstein-ij:Eq}) and (\ref{g:scalar eq:Eq}) give 
\Eqrsubl{g:cEinstein3:Eq}{
&&\hspace{-1cm}-(D-2)A''+\frac{c(D-2)^2}{2+c(D-2)}(A')^2+\frac{c}{4}K=0\,,
    \label{g:cEinstein3-yy:Eq}\\
&&\hspace{-1cm}R_{mn}(\Zsp)+\frac{c}{4}K w_{mn}(\Zsp)=0\,,
   \label{g:cEinstein3-mm:Eq}\\
&&\hspace{-1cm}b_0^{-2}\left[A''+(D-2)(A')^2\right]
+\frac{1}{4}\left(\frac{2}{D-2}+c\right)K=0\,,
   \label{g:scalar Eq3:Eq}
}
where $'$ denotes the ordinary derivative 
with respect to the coordinate $\theta$\,. Upon setting 
\Eqr{
c=-\frac{2}{D-1}\,,
    \label{g:ansatz for c:Eq}
}
we find 
\Eqr{
A(\theta)&=&\frac{1}{D-2}\ln\left[
\cos\left\{\sqrt{\frac{4\Lambda+(n-1)f^2}{2(D-1)}}
\left(
b_0\theta+b_1\right)\right\}\right]\nn\\
&&+b_2\,,
}
where $b_I~(I=0\,,1\,, 2)$ are constants. 
Hence, the metric of the $D$-dimensional spacetime can be written as
\Eqr{
ds^2&=&
\bar{A}^2
\left[\cos\left\{\sqrt{\frac{4\Lambda+(n-1)f^2}{2(D-1)}}
\left(
b_0\theta+b_1\right)\right\}
\right]^{2/(D-2)}\nn\\
&&\hspace{-0.4cm}\times
\left[q_{\mu\nu}(\Xsp)dx^{\mu}dx^{\nu}
+
b_0^2
d\theta^2
+w_{mn}(\Zsp)dz^mdz^n\right]\,,
  \label{g:D-metric3:Eq}
}
where $\bar{A}\equiv \e^{b_2}$ and $a_0$ are constants\,. 
We have 
obtained the de Sitter spacetime in a warped compactification,
under the general assumption
that both the field strengths
along the $\Xsp$ spacetime and the $\Ysp$ space
have nonvanishing values. 
But,
the field equations 
tell that
Ricci tensor on $\Xsp$ 
can take the positive sign,
even if either the field strength 
along 
the 
$\Xsp$ spacetime 
or that along the 
$\Ysp$ space
is vanishing.

We can 
rewrite the metric of the $D$-dimensional 
spacetime as 
\Eqr{
ds^2&=&
\bar{A}^2
\left[\cos\left(\bar{\theta}-\bar{\theta}_0\right)\right]^{2/(D-2)}
\left[q_{\mu\nu}(\Xsp)dx^{\mu}dx^{\nu}\right.\nn\\
&&\left.\hspace{-0.2cm}+\frac{2(D-1)}{4\Lambda+(n-1)f^2}\,d\bar{\theta}^2
+w_{mn}(\Zsp)dz^mdz^n\right],
  \label{g:D-metric4:Eq}
}
where 
$\bar{\theta}$ and $\bar{\theta}_0$ are defined by 
\Eq{
\bar{\theta}\equiv b_0\sqrt{\frac{4\Lambda+(n-1)f^2}{2(D-1)}}\,
\theta,
~~
{\bar\theta}_0
\equiv 
-b_1\sqrt{\frac{4\Lambda+(n-1)f^2}{2(D-1)}}\,.
}
This metric is of cohomogeneity one with 
foliations of two Einstein spaces
in $D$ dimensions
(for other examples of cohomogeneity one metric, see e.g.,
\cite{Cvetic:2001zb, Gibbons:2001ds, Kanno:2001xh, 
Cvetic:2001zx, Kanno:2001pg, Bilal:2001an, Cvetic:2001kp, 
Edelstein:2002zy, Cvetic:2005ft, Ahn:2005vc, Chen:2006xh, Chen:2006ea}).

Since the Kretschmann invariant of the metric (\ref{g:D-metric3:Eq}) 
is given by
\Eqr{
&&\hspace{-0.7cm}R_{ABCD}R^{ABCD}=\left[
\cos\left(\bar{\theta}-\bar{\theta}_0\right)
\right]^{\frac{-4}{D-2}}\left[n(n-2)\right.\nn\\
&&\left.+(D-n-1)(D-n-3)
+\frac{2(D-2)}{\cos^{4}\left(\bar{\theta}-\bar{\theta}_0\right)}\right],
  \label{div}
}
there are curvature singularities at 
\Eq{
\bar{\theta}=\bar{\theta}_0+\left(n+\frac{1}{2}\right)\pi 
\,,~~~~(n~{\rm is~integer})\,.
   \label{g:singularity2:Eq}
}
It is impossible to extend the spacetime across such a point
and we should restrict $\bar\theta$
to be for one period
$\bar{\theta}_0-\pi/2\le\bar{\theta}\le\bar{\theta}_0+\pi/2$.


\vspace{0.2cm}
%

We can construct very similar solutions
by considering only the 0-form $p_I=0$. 
Then the $D$-dimensional action in the Einstein 
frame includes  
the parameter of the 0-form field strength $m$, 
which is the dual
to the $D$-form field strength $F_{(D)}$ in the string frame. 
This is the equivalent to the contribution of the cosmological 
constant. 
If we set the coupling constant to the 0-form field strength
$\alpha_{(0)}=\alpha$, 
and use the ansatz for the metric 
(\ref{g:metric:Eq}) and 
for the scalar field (\ref{g:ansatz of scalar:Eq})
with (\ref{g:c1:Eq}), we can obtain the de Sitter solutions which have 
very similar metric form to (\ref{g:D-metric4:Eq}).

\vspace{0.2cm}

In summary,
we have presented 
exact solutions of
the external $n$-dimensional de Sitter spacetime
with the 
compactifications on the $(D-n)$-dimensional compact Einstein space 
under certain conditions
in the $D$-dimensional gravitational theory coupled to the dilaton,
$n$- and $(D-n)$-form field strengths 
and a cosmological constant (\ref{g:action:Eq}).
We also 
commented 
in the theory with the 0-form field strength.
The metric structure requires that warp factor is expressed as the form 
$\e^{A(y)}$, where $A(y)$ is a function on the internal space. 
We could find the de Sitter compactification if the matter fields with 
background parameters 
satisfy the equations (\ref{g:Lambda:Eq}).
The $n$-dimensional spacetime becomes de Sitter while the internal space has 
a finite volume. 
Then, it is possible to find 
a de Sitter compactification with a negative 
cosmological constant.

This work is still at the starting point
for the research in this direction.
There are problems which should be resolved.
In our analysis, 
we could not find de Sitter solutions in 
the ten- or eleven-dimensional supergravity because 
the scalar 
coupling constants are 
different from those
in 
it. If we choose the couplings in the supergravity,  
we only obtain
the negatively curved $n$-dimensional spacetime 
and hence AdS spacetime. 
There 
 were several trials 
to find the de Sitter solution 
in the supergravity theory in terms of the dynamical $p$-brane solutions. 
However, it 
was impossible to find the de Sitter solution 
in  the 
ten- or eleven-dimensional supergravity theory \cite{Gibbons:2005rt, 
Chen:2005jp, Kodama:2005fz, Kodama:2005cz, Binetruy:2007tu, Maeda:2009tq, 
Maeda:2009zi, Gibbons:2009dr, Maeda:2009ds, Minamitsuji:2010kb, 
Maeda:2010yk,Maeda:2010ja, Minamitsuji:2010fp, Maeda:2010aj, 
Minamitsuji:2010uz, Minamitsuji:2011jt}
if  we assume the scalar coupling constants
in 
 supergravity \cite{Maki:1992tq, Minamitsuji:2010uz}.

We also assumed the simple internal space as an example
 as S$^1\times$S$^{D-n-1}$ . 
However, 
for instance, 
if the internal space is assumed to be T${}^2\times$T${}^{D-n-2}$,
the $n$-dimensional spacetime should be flat or
a negative curvature spacetime.
Hence, it is not so easy to find de Sitter compactification in the Calabi-Yau 
internal space. 
Among more recent developments, de Sitter solutions in string theory
 have been found to have a connection to the four-dimensional inflationary 
 scenario with extended supersymmetry and 
 have emerged as an
important component of the internal spaces with enough symmetries of 
de Sitter compactification. 
In our analysis, we have focused on the availability of solutions 
for a given ansatz,
and will leave the issues of the stabilization
and the cure of curvature singularities
for future studies.
The next step 
of our work 
will be to develop a framework to understand 
string compactifications in which one varies the matter fields, 
the boundary conditions or other details.
The issues on the stabilization
and the cure of curvature singularities
are also very crucial for a realistic model of the dark energy.
All of them should be clarified in our future studies. 

\section*{Acknowledgments}
The work of M.M. was supported by the Yukawa fellowship.
K.U. is supported by Grant-in-Aid for 
Young Scientists (B) of JSPS Research, under Contract No. 20740147.



\end{document}